# Nanoscale assembly processes revealed in the nacroprismatic transition zone of *Pinna nobilis* mollusc shells

Robert Hovden[1,*], Stephan E. Wolf[2,3,*], Megan E. Holtz[1], Frédéric Marin[4], David A. Muller[1,5] & Lara A. Estroff[2,5]

Intricate biomineralization processes in molluscs engineer hierarchical structures with meso-, nano- and atomic architectures that give the final composite material exceptional mechanical strength and optical iridescence on the macroscale. This multiscale biological assembly inspires new synthetic routes to complex materials. Our investigation of the prism–nacre interface reveals nanoscale details governing the onset of nacre formation using high-resolution scanning transmission electron microscopy. A wedge-polishing technique provides unprecedented, large-area specimens required to span the entire interface. Within this region, we find a transition from nanofibrillar aggregation to irregular early-nacre layers, to well-ordered mature nacre suggesting the assembly process is driven by aggregation of nanoparticles (~50–80 nm) within an organic matrix that arrange in fibre-like polycrystalline configurations. The particle number increases successively and, when critical packing is reached, they merge into early-nacre platelets. These results give new insights into nacre formation and particle-accretion mechanisms that may be common to many calcareous biominerals.

[1] School of Applied and Engineering Physics, Cornell University, Ithaca, New York 14853, USA. [2] Department of Materials Science and Engineering, Cornell University, Ithaca, New York 14853, USA. [3] Department of Materials Science and Engineering, Institute of Glass and Ceramics, Friedrich-Alexander-University Erlangen-Nürnberg, 91058 Erlangen, Germany. [4] UMR CNRS 6282 Biogéosciences, Université de Bourgogne Franche-Comté, 6 Boulevard Gabriel, 21000 Dijon, France. [5] Kavli Institute at Cornell for Nanoscale Science, Ithaca, New York 14853, USA. * These authors contributed equally to this work. Correspondence and requests for materials should be addressed to L.A.E. (email: lae37@cornell.edu).

Comprised of calcium carbonate ($CaCO_3$) polymorphs—primarily aragonite and calcite—mollusc shells span a variety of structures[1–3]. Their superior hardness, strength and toughness inspire the artificial synthesis of biomimetic materials[4–8] and motivate understanding the biological mechanisms of formation[9,10]. The most familiar mollusc-derived biomineral is the lustrous nacre (that is, mother of pearl) found in the interior of the animal's shell[11–13]. The 'brick-and-mortar' mesostructure of nacre is a stacked arrangement of polygonal aragonite tablets with a thickness comparable to the wavelength of light (~500 nm), giving rise to its iridescence[14–16]. In bivalves, for example, in the order of *Pterioida*, the nacre often grows atop an outer prismatic layer, an assembled array of elongated prismatic calcite crystals aligned perpendicular to the shell's surface.

The transition from the outer shell's prismatic calcite to the nacreous aragonite interior has remained perplexing. In nacroprismatic shells, a thick organic layer (sometimes referred to as the 'green sheet') is present at the prism-to-nacre transition of molluscs and is hypothesized to play a key role in directing the switch from the growth of prismatic calcite to platelet aragonite[10,17]. The nanoscale structure of the $CaCO_3$ assembly occurring throughout this organic layer has, however, not been examined in detail. Previous work has suggested the presence of a disordered aragonite layer at the prismatic–nacre transition zone of both gastropods and bivalves[17–19]. For example, Dauphin *et al.*[18] identified granular material, which they termed 'fibrous aragonite', across the prism–nacre interface of the bivalve *Pinctada margaritifera*. Until now, however, the complete nanoscale morphology of the nacroprismatic transition has not been reported. Employing high-resolution scanning transmission electron microscopy (STEM) and sample preparation techniques popular in the semiconductor industry (see Methods), we are now able to reveal the nanoscale details of the mineralization process occurring throughout this transition.

The Noble Pen Shell *Pinna nobilis* (*Pterioida*, Linnaeus 1758), endemic to the Mediterranean, is an established model system for bivalve shell formation[20–27] and served as a representative nacroprismatic shell structure for our investigation. As the mollusc shell is deposited by the mantle cells on the periostracum[2,28,29]—a thin organic membrane separating the shell from the surrounding marine environment—the shell is built 'from the outside in'. In addition to mineral components, the epithelial cells secrete extracellular matrix components that provide both the mesoscale structural framework and control of the mineralization process. This organic matrix is composed mainly of various proteins, some of which are highly acidic or glycosylated, and polysaccharides, for example, β-chitin[30]. These components are thought to template and regulate the self-organization process on the micro- to nanoscale during biomineral formation[9,10,31–33].

The switch from prismatically organized calcite to densely stacked aragonite tablets, that is, nacre[11,34–36], is orchestrated by finely tuned secretion of organic and inorganic ingredients into an extracellular compartment—the so-called extrapallial space. The complex interplay between the organic and inorganic constituents then governs the formation of the mineralized shell matrix. It was shown recently that the switch from the prismatic to the nacreous shell layer in *P. margaritifera* and *P. maxima* is accompanied by a marked switch of the secretory repertoires; *in situ* hybridization evidenced a remarkably sharp transition from the prism to the nacre transcript expression in the epithelial cells[31,37,38]. Chemical modifications at the end of prisms, before the formation of the transition organics has also been observed[18,36]. It should be noted that both nacreous and prismatic epithelial mantle cells secrete their constituents into the same extrapallial space—there is no membrane separating the two regions.

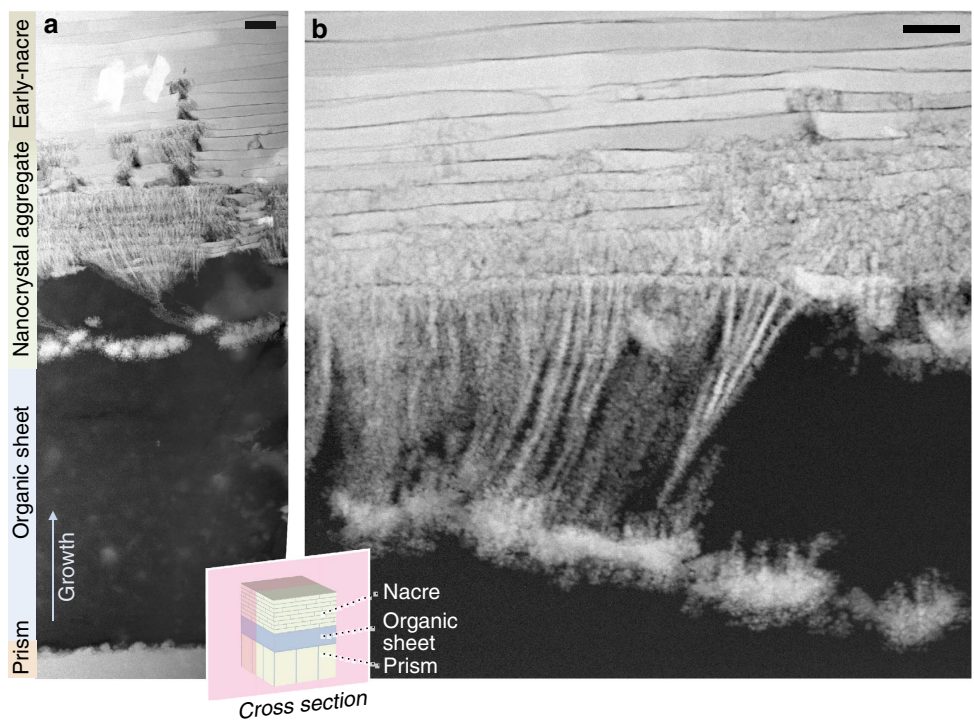

**Figure 1 | The start of nacre growth at the nanoscale imaged with ADF-STEM in cross-section across the prismatic to nacreous transition.** Prismatic growth is abruptly terminated with a thick organic sheet (**a**, bottom). Several microns later, $CaCO_3$ growth resumes (**a**, middle), first initiated with a layer of larger nanoparticles (~700 nm), then followed by deposition and progressive aggregation of small nanoparticles—more clearly seen at higher magnification of another region in **b**. The nanoparticles form a branching fibrillar structure with increasing aggregation of particles. These structures then lead to the onset of disordered nacre layers (early nacre). Scale bars, (**a**) 1 µm; (**b**) 500 nm.

Previous investigations have proposed that shell growth in adult molluscs proceeds via a nanoparticle-accretion mechanism in which amorphous nanoparticles are deposited, aggregate and crystallize within the extrapallial space[22,39]. Similarly, in bivalve larvae, vibrational spectroscopy has shown amorphous calcium carbonate that later transforms into crystalline aragonite[40] and traces of amorphous calcium carbonate can be found even in adult specimens[22,29,33]. Furthermore, *in vitro* evidence for a particle-accretion mechanism has been suggested for numerous other calcarous biominerals[41]. Definitive identification of metastable phases, however, is notoriously difficult as it is complicated by both the transient nature of the metastable phase[3,29,33,42–45] and the similarity of the diffraction pattern to nanocrystalline particles[46]. Recently, *in situ* transmission electron microscopy experiments have observed the nucleation and growth of amorphous CaCO$_3$ in a matrix of polystyrene sulphonate[47]. Direct evidence of similar processes *in vivo* requires sample preparation and imaging techniques with the capability to reveal nanoscale details.

In this work, we investigate the nacroprismatic transition zone to provide insight into the delicate processes of self-organization by which bivalves control the switch from calcitic prisms to aragonite tablets. Imaging this nacroprismatic transition zone with sub-nanometre resolution, we find a layer of nanocrystallites in fibre-like arrangements within a thick organic layer preceding nacre. This nanofibrillar aragonite branches upward (that is, in the direction of shell growth), with increasing coverage and density in each consecutive growth layer until it finally forms continuous aragonite nacre stacks by the merging and fusion of individual calcium carbonate particles to a dense, space-filling nacre platelet. These early-nacre layers are structurally distinct from mature nacre—with irregular layer thickness, greater layer interface curvature, higher polycrystallinity and frequent stacking faults. The entire transition from prismatic calcite to nacre can occur over tens of microns; however, the fundamental nano-sized building units are only 50 nm and not directly observable by standard optical or X-ray techniques. High-resolution STEM allows us unique, sub-nanometre resolution visualization of the processes of the switch from prismatic to nacre formation. This work provides direct evidence for a particle-assembly pathway occurring during this transition, which is consistent with the model proposed by Gal *et al.*[41,48,49]

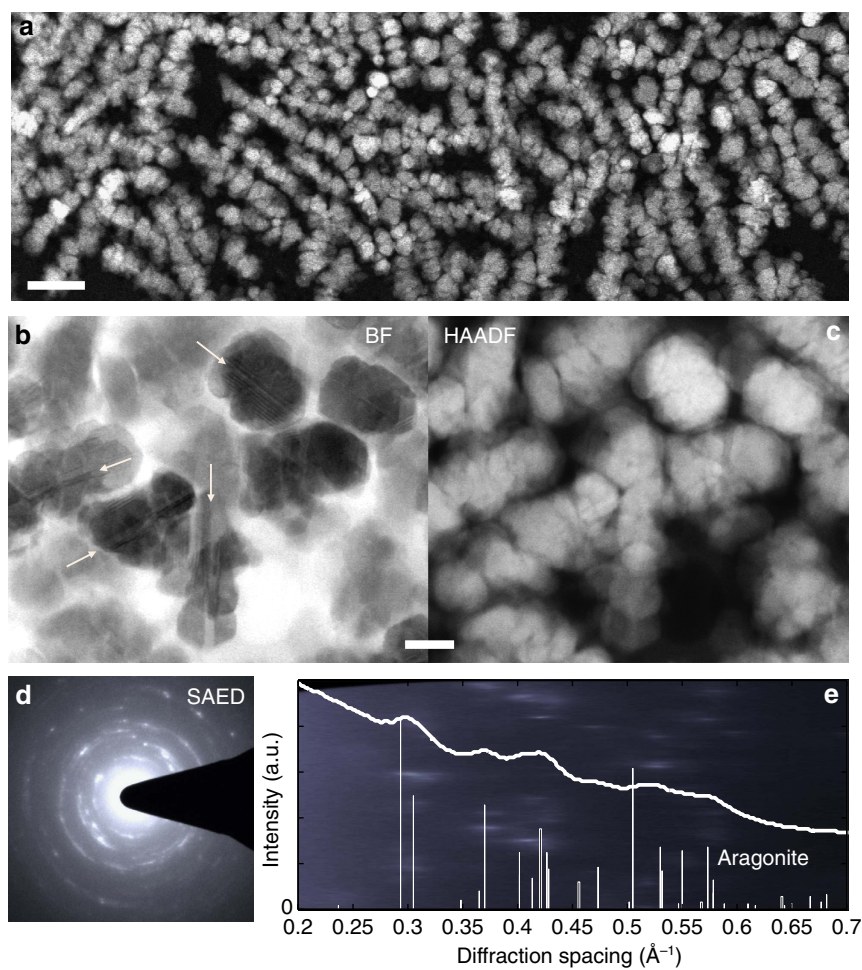

**Figure 2 | Structure of the pre-nacre nanocrystals showing a typical diameter of 50–80 nm.** (**a**) Planar view of the packing in pre-nacre aggregation region. (**b**,**c**) Pair of annular dark-field (HAADF) and bright-field (BF) STEM images taken in the same region. The BF image of nanocrystals clearly reveals polycrystallinity among the particles and within each particle; twin boundaries running end to end are highlighted with arrows. (**d**) SAED shows that the nanocrystallites are consistent with the aragonite CaCO$_3$ polymorph—the same polymorph of mature nacre. (**e**) Plot of radially integrated SAED intensity with known aragonite peaks marked along the x axis; background of plot **e** is a polar transformation of **d** over 0–5 radians. Scale bars, (**a**) 200 nm; (**b**,**c**) 50 nm.

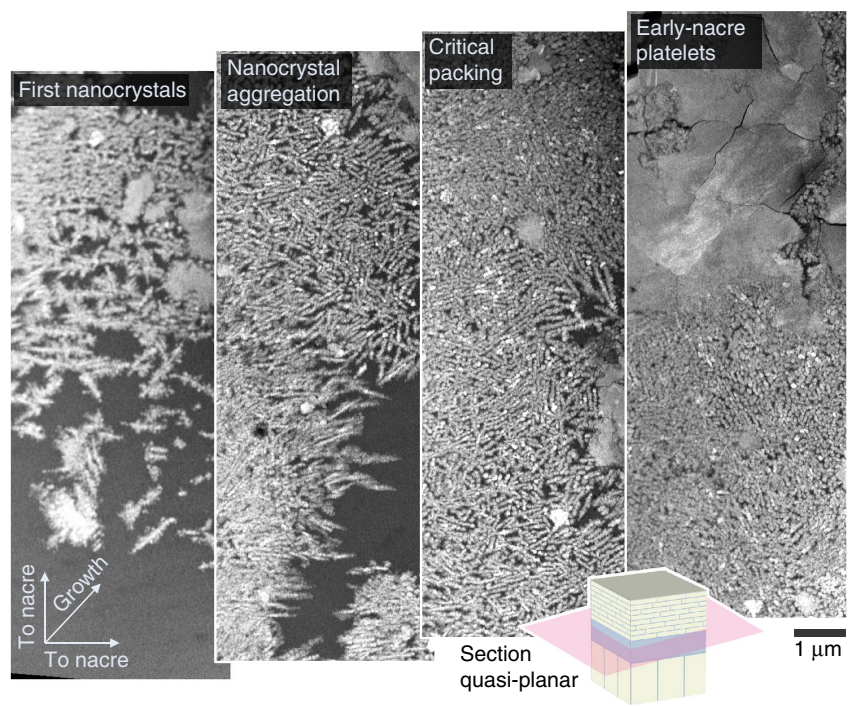

**Figure 3 | Quasi-planar nanoscale view at the onset of mollusc nacre growth shown by HAADF-STEM.** Nacre growth begins by aggregation of nanocrystals in the organic matrix that separates the prismatic from the nacreous growth (lower left). As growth progresses (towards top right), the nanocrystals aggregate and grow with disordered, often needle-like, geometries. When a critical packing density is reached, continuous early-nacre platelets form—later leading to the familiar large, uniform nacre plates. The quasi-planar section geometry is illustrated in bottom inset.

## Results

**Observation of the prism-to-nacre transition.** Biomineralization at the mollusc's prism-to-nacre transition is governed by nanoscale mechanisms visible to high-resolution STEM. However, the entire prism–nacre interface spans tens of microns, which represents a clear challenge for sample preparation. Pertinent sample preparation methods, such as ion milling and thinning approaches, only provide small areas of a few microns suitable for STEM investigation. Furthermore, biominerals represent composite materials consisting of calcium carbonate and organics that are very sensitive to irradiation damage by electrons or ions. We thus implemented a wedge-polishing technique adapted from the semiconductor industry for sample preparation of remarkably large electron-transparent regions of excellent quality and without the use of ion milling techniques. Several samples from the nacre–prism interface in *P. nobilis* were prepared with this technique (Supplementary Fig. 1). Imaging of these samples demonstrates the power of a wedge-polishing approach: the entire nacroprismatic transition is captured in a single image from the prismatic (Fig. 1a, bottom) to nacreous (top) layer.

Since the bivalve shell is formed from layer-by-layer deposition, the cross-section illustrates sequential nanoscale processes at the onset of early-nacre formation. Early nacre here refers to the first nacre that is formed in the nacroprismatic transition, which actually makes it the oldest nacre in the shell. We can distinguish three stages in the spatial transition from the prismatic to the nacreous layer:

1. Abrupt prism termination: a continuous layer of organic material marks a relatively abrupt termination of prismatic structure. From annular dark-field (ADF)-STEM images, this organic layer has a density indistinguishable from the intra-prismatic organic, appearing as a continuation of the same or similar material (Supplementary Fig. 2). In Fig. 1a, complete coverage of roughly 10-µm-thick organic material was observed before the restart of any significant calcium carbonate deposition.

2. Nanoparticle aggregation: within this organic buffer layer separating the prismatic and the nacreous layers, nacre is preceded by the formation of $CaCO_3$ nanoparticles (Supplementary Fig. 3) that tend to aggregate in a fibre-like geometry. Following initial deposition of these nanoparticles, the particle number increases as the shell growth continues— creating a divergence of densely packed particulate material (Fig. 1) that leads to complete coverage of fibrous aggregates. This region still lacks the mineral density of nacre but often exhibits periodic reduction in density occurring with frequency comparable to the spacings of early nacre. Eventually, the packing density of these aggregates increases until the crystallites merge to form continuous nacre platelets (Fig. 1a,b, top).

3. Formation of early-nacre platelets: the onset of nacre formation is generally incomplete, with early-nacre platelets neighbouring high-density nanoparticle aggregates (Fig. 1, early nacre; and Supplementary Figs 4 and 5). As growth continues, more highly packed aggregate regions become continuous nacre plates. In some instances, nacreous growth returns to aggregate particles. Eventually, the growth forms complete continuous nacre layers (Fig. 1a, top). These early-nacre layers are similar to mature nacre but more disordered at the mesoscale (for example, variation in layer thickness, crystal orientation, and containing stacking faults).

**Nanocrystal structure.** Closer inspection of the nanoparticles reveals polycrystalline aragonite ~50 nm in diameter (Fig. 2). The polycrystallinity is most visible in bright-field STEM images (Fig. 2b)—where contrast is sensitive to changes in crystallographic orientation and structural defects. Observation by selected-area electron diffraction (SAED) confirms crystallinity throughout the nanocrystals and finds them to be consistent with the aragonite polymorph (Fig. 2d,e and Supplementary Fig. 6). Aragonite is also the polymorph found in mature nacre. No amorphous calcium carbonate (ACC) nanoparticles were observed from casual observation via convergent beam electron diffraction nor were they suggested by the bright-field STEM images. As discussed in the Methods section, great care was taken to ensure that no phase transformation occurred during sample preparation and imaging. As such, the observed phase and structure of these particles represents their state at the time the 60-cm mollusc specimen was collected. However, based on this data, there is no way of deducing what the original phase of the particles was at the time of deposition or tracking any phase transformations that may have occurred. Consistent with recent literature regarding the role of ACC in the biomineralization of mollusc shells[3,12,22,29,41] it is likely that the particles were initially ACC and subsequently crystallized, either directly after deposition or during the ageing of the shell.

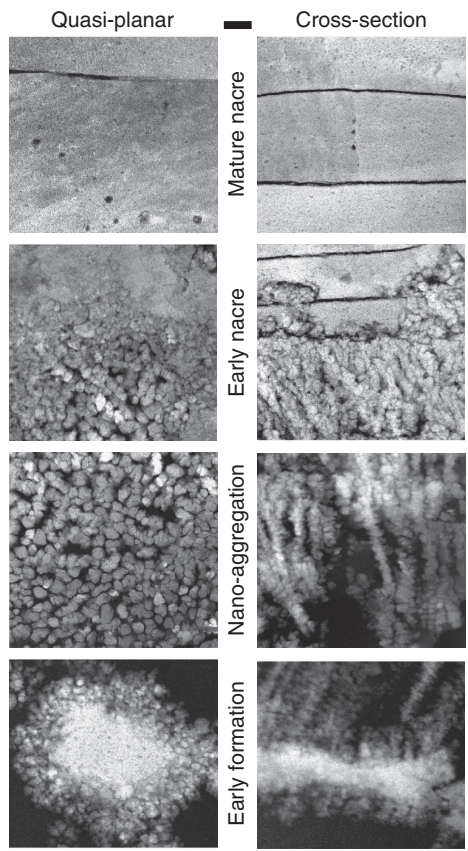

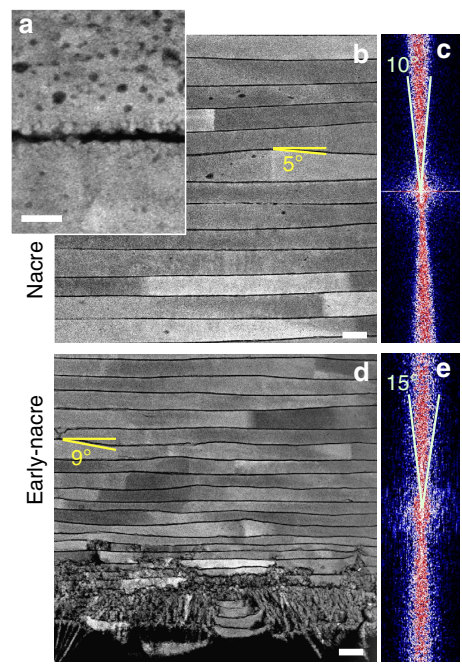

**Figure 4 | HAADF-STEM images of nacre formation viewed from orthogonal directions.** Early formation begins with the appearance of CaCO$_3$ layers with planar disk geometry (>1 μm diameter, ~300 nm thick), after which nano-aggregation of ~60 nm crystals of aragonite accumulate along the growth direction with increasing density. Once a critical packing density is reached, early-nacre platelets form as a packed continuum with adjacent nanocrystals. Nacre plates are separated by a thin (~8 nm) organic sheet. Eventually, uniform ordered nacre platelets are formed. Organic spheroid inclusions can be seen in nacre. Scale bar, 200 nm.

**In-plane observation of the assembly.** The structure and packing density of the pre-nacre nanocrystals (nanofibrillar aragonite) is more clearly seen in the plane of the growth. To complement cross-sectional investigation, subsequent planar sections, perpendicular to the growth direction, were prepared. Adding a small planar tilt (that is, quasi-planar sectioning) allowed in-plane observation throughout the entire prism–nacre interface. The in-plane aggregation process throughout the prism–nacre transition is shown in Fig. 3—beginning in the bottom left-most image and advancing with growth towards the top right-most image (Supplementary Fig. 7). Following the initial appearance of nanofibrillar aragonite within the organic matrix (Fig. 3, first nanocrystals) the growth progresses by increased planar packing of these nanocrystals. The nanocrystals are often connected with a fibrous, or needle-like, geometry. This arrangement may be due to in-plane columnar growth processes or due to a template action of chitin fibres[9,39]; clarification of the underlying processes will be the subject of future research.

**Figure 5 | Comparison of early-nacre and mature nacre growth shown by HAADF-STEM.** (**a**) Layers have well-defined interfaces buffered by ~8-nm organic layers. For mature nacre (**b**), the layers are ordered and have interface curvature ranging ±5.0° as clearly shown by the image's Fourier transform (**c**). The first several nacre layers, that is, early nacre (**d**), are more disordered and have an interface curvature ranging roughly ±7.5° as shown by the image's Fourier transform (**e**). As seen in **d**, the interlayer spacing in early nacre increases progressively—starting as small as 120 nm. Mature nacre (**b**) has a spacing of ~450 nm. Scale bars, (**a**) 50 nm; (**b**,**d**) 500 nm.

The early-nacre platelets appear to be formed by the fusion of individual nanoparticles when a critical particle number density is reached (Fig. 3, top right). The transition from packed nanocrystals to nacre is a continuous process. At the rim of the forming platelets we see nanogranular features and the gradual transitioning from individual nanocrystallites to space-filling nacre platelets (Supplementary Fig. 4). Variation of high-angle annular dark-field (HAADF) image intensity in the early-nacre platelets reflects their polycrystallinity.

**Comparison of cross-sectional and quasi-planar views.** The three-dimensional structure at the transition can be inferred by comparing the samples that were polished both parallel and perpendicular to the growth direction. Figure 4 compares the two sectioning geometries for high-resolution STEM imaging at the three observed stages in the transition zone. The earliest CaCO$_3$ formations (Fig. 4 early formation) appear with planar disk geometry (>1 μm diameter, ~300 nm thick) and have SAED peaks consistent with aragonite (Supplementary Fig. 8). In some areas, nanofibrilliar aragonite originates from these initial structures leading to the onset of nacre, whereas in other locations, these disk-like structures appear isolated in cross-section (Fig. 1). The nano-aggregation then proceeds: ~60 nm aragonite crystals accumulate along the growth direction with increasing density (Fig. 4 nano-aggregation). These nanocrystals are often linearly connected along the growth or in-plane directions. The early formations often—but not always—provide a site of growth for these nanocrystals (as seen in Fig. 1a). Once a critical nanocrystal packing density is reached, continuous early-nacre platelets form adjacent to packed nanocrystals (Fig. 4, early nacre). Eventually, uniform ordered nacre platelets are formed (Fig. 4, mature nacre). Organic spheroid inclusions can be seen in the nacre tablets as previously reported[50].

**Structure of early-nacre platelets.** Additional structural variation in early-nacre makes it distinguishable from mature nacre, which is characterized by uniform tablets (Fig. 5a–c and Supplementary Fig. 9). Following the first appearance of nacre platelets, subsequent layers contain larger continuous regions, until complete early-nacre layers are formed. However, the early-nacre layers (Fig. 5d) are not structurally equivalent to mature nacre (Fig. 5b). The thin organic layers (~10 nm or less) separating all nacre layers (Fig. 5a) provide a well-defined, high-contrast boundary for characterizing the planar variation via real and Fourier space analysis. Mature nacre consists of ordered platelets roughly 450-nm thick with small, slow varying changes in thickness. The Fourier transform reveals a planar variation tightly bound within ±5° from the growth plane (Fig. 5c). In comparison, early-nacre layers are thin (as low as ~120 nm) and disordered. Rapid variation in layer thickness and interface curvature is clearly seen in the cross-sectional image (Fig. 5d) and reflected in the wide angular intensity in Fourier space (Fig. 5e). We observe a planar variation ca. ±7.5° (Fig. 5e) in the early nacre. Changes in HAADF-STEM diffraction contrast, within a single nacre tablet, can be seen in the bright-to-dark variation. This observation corresponds with small variation of the in-plane crystal orientation of nacre plates—crystallographic changes that were previously reported by Gilbert et al.[12,51]. We additionally observe that changes in crystal orientation are more prevalent in the early-nacre when compared with mature nacre. We confirm, with high resolution, the presence of inorganic bridges between nacre layers in the bivalve P. nobilis, as, for instance, previously reported by Checa et al.[52]. High-resolution micrographs of the bridges are shown in Fig. 5a and Supplementary Fig. 10.

## Discussion

In this work, P. nobilis samples were prepared by a wedge-polishing technique for sub-nanometre observation of the nacroprismatic transition zone with high-resolution ADF-STEM and SAED. This contribution illuminates the formation of the first layers of nacre and demonstrates that nacre onset is a complex nanoscale assembly process. The onset of nacre growth occurs from the aggregation of nanoparticles ca. 50–80 nm in size which then pack as polycrystalline fibre-like arrangements (nanofibrillar aragonite), branching outward with increasing density in the growth direction. When the particle number density reaches a critical value, the particles merge and form the first early-nacre platelets. Thus, the highly ordered state of nacre is achieved gradually, transitioning from the densely packed nanofibrillar aragonite, to an early-nacre layer with a higher rate of disorder and eventually, the well-ordered mature nacre layers.

The cross-section of the nacroprismatic transition zone is a representation, frozen in time, of the transition from one mineralization mechanism to another. In other words, using the current techniques, the observed ultrastructure is well preserved and representative of the in vivo state (at harvest) and provides insight into the assembly of early nacre. In this way, the observations reported here might be analogous in some respects to the formation process of mature nacre tablets, even in layers further away from the transition zone. We cannot, however, see back in time to the dynamics or material phases that occurred in the tissue during the nanoparticle formation and assembly—including the possible crystallization of ACC. These findings give new insights into the self-organization mechanisms of nacre that may also be common to many calcareous biominerals.

## Methods

**Specimen.** The protected Mediterranean P. nobilis specimen was collected near the coast of Villefranche-s-Mer, Département Alpes-Maritimes, France with the authorization of DREAL PACA (Direction Régionale de l'Environnement, de l'Aménagement et du Logement, Provence Alpes-Côte d'Azur).

**Wedge-polishing preparation.** Small sections were isolated from the 60-cm shell (Supplementary Fig. 1) using a diamond wire saw. Specimen preparation for STEM analysis required undamaged, electron-transparent regions over large areas capable of spanning the entire prism–nacre interface—both the $CaCO_3$ and organic material are easily damaged by electron and ion beams. We successfully implemented a wedge-polishing technique that is particularly popular in the semiconductor industry[53]. The polishing technique is gentle and structural changes have not been observed by this method. Dislocated material or granules did not occur—however would be easily noticed in a STEM image as a vacancy and followed by a damage path caused by the removed material during polishing. The lapping process was conducted with an Allied MultiPrep polishing system and used a waning series of micron-sized diamonds embedded in polishing film (from 30 μm down to 0.1 μm) dictated by the 'trinity of damage': polishing produces a damaging layer, which is approximately three times thick as the size of the grit. Therefore, with each subsequent lapping film, this damage layer has to be removed. To prevent $CaCO_3$ etching during preparation, water-based lubricants and polishing agents were eliminated—instead, Allied alcohol based 'blue' lubricant was used. For the first-side polish, the sample was attached to an aluminum-polishing stub (Allied) with Crystal Bond (Allied) and was polished until a polish free of scratches and imperfections is obtained at the desired location. For the second-side polish, the sample was glued to a Pyrex-stub (Allied) with a very thin layer of superglue. The polishing procedure was essentially repeated, but at a 3° pitch to form a wedge. The orientation, pitch and roll of the sample were additionally adjusted with two micrometer heads so that the sample plane of interest was parallel to the abrasive plane. Finally, the polished wedge samples were mounted to an annular molybdenum transmission electron microscopy grid with MBond 610 epoxy. By optimizing this polishing technique for calcium carbonate biominerals, we could obtain samples with areas up to 3-mm wide, which were accessible to STEM analyses and of high quality.

**Electron microscopy.** STEM was performed using a 200-keV FEI F20 instrument with a convergence angle ca. 9.6 mrad. A HAADF with 100–300 mm camera lengths was used to provide a Z-contrast image where intensity is sensitive to the atomic number of atoms in the specimen. A bright-field detector with a 200-mm camera length was used to highlight polycrystallinity in the specimen. SAED with an aperture of approximated 8 μm provided a polycrystalline diffraction pattern of nanocrystallites in the prism–nacre transition regions.

### Acknowledgements
L.A.E. acknowledges support from the NSF (DMR-1210304). This work and support for R.H. made use of the Cornell Center for Materials Research Facilities supported by the National Science Foundation under Award Number DMR-1120296. We thank John Grazul and Mick Thomas for experimental assistance, and Sébastien Motreuil (UMR CNRS 6282 Biogéosciences) who sampled *Pinna nobilis*. S.E.W. thanks for partial support by the Cluster of Excellence 315 'Engineering of Advanced Materials—Hierarchical Structure Formation for Functional Devices' funded by the German Research Foundation.

### Author contributions
R.H. and S.E.W. contributed equally to this work; sample preparation was developed by R.H., S.E.W. and M.E.H.; electron microscopy was conducted by R.H. and S.E.W.; data analysis and materials interpretation was carried out by R.H., M.E.H. and D.A.M.; biological context was provided by S.E.W. and L.A.E.; protected *Pinna nobilis* specimen and valuable input was provided by F.M.; all authors discussed the results throughout all stages and commented on the manuscript. R.H., S.E.W. and L.A.E. wrote the manuscript.

### Additional information
**Supplementary Information** accompanies this paper at http://www.nature.com/naturecommunications

**Competing financial interests:** The authors declare no competing financial interests.

**Reprints and permission** information is available online at http://npg.nature.com/reprintsandpermissions/

**How to cite this article:** Hovden, R. *et al.* Nanoscale assembly processes revealed in the nacroprismatic transition zone of *Pinna nobilis* mollusc shells. *Nat. Commun.* 6:10097 doi: 10.1038/ncomms10097 (2015).